\documentclass[11pt]{article}
\usepackage[utf8]{inputenc}
\usepackage{amsmath,amsfonts,amssymb,latexsym}
\usepackage[T1]{fontenc}
\newtheorem{satz}{Theorem}[section]

\newtheorem{defi}[satz]{Definition}

\newtheorem{bem}[satz]{Remark}

\newtheorem{koro}[satz]{Corollary}
\newtheorem{conclusion}[satz]{Conclusion}
\newtheorem{ob}[satz]{Observation}

\newcommand{\mcal}{\mathcal}

\newcommand{\tit}{\textit}

\newcommand{\R}{\mathbb{R}}

\newcommand{\beq}{\begin{equation}}
\newcommand{\eeq}{\end{equation}}

\begin{document}
\thispagestyle{empty}
\begin{center}
\vspace*{1.0cm}
{\Large{\bf On the Reality of Unruh Temperature}}
\vskip 1.5cm

{\large{\bf Manfred Requardt}}

\vskip 0.5cm

Institut fuer Theoretische Physik\\
Universitaet Goettingen\\
Friedrich-Hund-Platz 1\\
37077 Goettingen \quad Germany\\
(E-mail: requardt@theorie.physik.uni-goettingen.de) 

\end{center}
 
\begin{abstract}
In contrast to recent criticism we undertake to show that the notion of Unruh temperature describes a real thermal property of the vacuum if viewed from an accelerated reference frame. We embed our investigation in a more general analysis of general relativistic temperature (Tolman-Ehrenfest effect) with the entropy-maximum principle being our guiding principle. We show that the Unruh effect neatly fits into this more general framework. Our criterion of reality is, first, the possibility to transfer a quantum of acceleration radiation to the inertial laboratory where it can be studied in principle under ordinary thermodynamical conditions. Second, we emphasize as another criterion the importance of the coincidence of the accelerated and inertial observer description as far as the final objective result is concerned.
\end{abstract} \newpage
\setcounter{page}{1}
\section{Introduction}
Recently Buchholz and Solveen (\cite{Bu}) criticized the standard interpretation of the Unruh-effect, i.e. that the accelerated observer sees and/or detects a gas of so-called \tit{Rindler-particles} of temperature $a/2\pi$ with $a$ being the \tit{proper acceleration} on his respective Killing orbit. The authors argued that the notion and concept of \tit{temperature} is usually used too carelessly in this context and that for an accelerated observer the vacuum appears as cold as for an inertial observer if temperature is interpreted in a certain way, which the authors furthermore argue is the correct way. They claim that a certain occurring parameter is rather and only a measure of acceleration and not of thermal behavior. As their arguments are carefully laid out and their claims are not the standard ones, it is necessary in our view to rethink the whole field in a systematic way and to scrutinize several of the commonly used concepts. 

In our investigation we will rely on two seminal papers by Unruh and Wald (\cite{Wa1},\cite{Wa2}), which show among other things how intricate an exhaustive analysis has to be in order that every point and physical aspect is clarified.
\begin{ob} In order to fully understand the phenomena in this field it is crucial to constantly employ two points of view or reference frames, the framework used by the accelerated observer and the point of view adopted by the inertial observer. Only if these two observers are able to describe the same physical phenomena coherently in their (in general) different languages, can we be certain that the phenomena are real and not only fata morganas.
\end{ob}
\begin{bem} Note that this principle is also extremely fruitful in black hole physics (BH) and is called there BH-complementarity (cf. the discussion in \cite{Susskind1},\cite{Susskind2}. 
\end{bem}
Ultimately it is the philosophy behind large parts of general relativity (GR). That is, objective phenomena occurring in the event manifold, described differently but in a complementary way in different coordinate systems or reference frames. A beautiful (paradigmatic) example (the accelerating rockets, connected by a thread) can be found in the essay by Bell (\cite{Bell}). We shall stick to this philosophy in the following sections.

In the next section we introduce some temperature concepts and exhibit the simple but important relation between the time parameter and the temperature scale in a given setup. It underlies in fact the whole field of relativistic thermodynamics and in particular the Unruh effect. In section 3 we discuss thermometer observables and emphasize the necessity of observables which are sensitive to acceleration and not only instantaneous velocity. In section 4 we use the \tit{entropy-maximum principle} in order to introduce the notion of local relativistic temperature. We note that, in contrast to the non-relativistic regime, a temperature gradient is compatible with thermodynamic equilibrium. We discuss in quite some detail the so-called \tit{Tolman-Ehrenfest effect}.

In section 5 we introduce the \tit{relativistic Carnot cycle}. In section 6 we review the analysis by Unruh and Wald of the Unruh effect. It is our point of view that the reality of e.g. Unruh temperature is proved if one is able to extract a quantum of heat energy from the acceleration regime and transfer it to the laboratory. This however follows from the \tit{mining process} described by the authors. From conservation of energy it follows that the extracted heat energy is a transmuted form of acceleration work which has to be supplied by the external source which accelerates observer, system or whatever and which is not part of the microscopic system itself.

\section{Some Remarks on Generalized Temperature Concepts}
We begin with some notations employed in \cite{Wa1}. Unruh/Wald use a Killing vector field in the right wedge, $W_R$, of the form
\beq b^{\nu}:=a(x\cdot t^{\nu}+t\cdot x^{\nu})\quad\text{(abstract index notation)}  \eeq
with an arbitrary constant $a$ and
\beq t^{\nu}:=\partial/\partial t\quad ,\quad x^{\nu}:=\partial/\partial x   \eeq
\begin{bem} If we make the standard choice
\beq t=x^0=\xi\sinh\eta\quad ,\quad x=x^1=\xi\cosh\eta    \eeq
with $\eta$ the Rindler time , we have $a=1$.
\end{bem} 
The calculations in \cite{Wa2} then show that the constant $a$ corresponds to the proper acceleration on the orbit along which 
\beq  -b^{\nu}b_{\nu}\equiv 1\quad\text{(convention -+++)}   \eeq
Along this orbit Killing parameter time equals proper time. With this choice the temperature in $W_R$ is 
\beq  T=a/2\pi  \eeq
(cf. also \cite{Wa3}). It is the local temperature an observer measures along the orbit with proper acceleration $a$.

If, on the other hand, we approach this field from another direction, i.e. view $W_R$ as a subregion of Minkowski space $M$ and then, via the \tit{Tomita-Takesaki-KMS theory}, regard the global pure Minkowski vacuum state restricted to $W_R$ as the corresponding KMS state, we arrive at an abstract KMS-temperature (depending on the normalization) like, for example, $2\pi$ (cf. \cite{Haag} or \cite{Wi}) in case we associate the boost generator with the Tomita Hamiltonian. However if we place these observations in a wider context we learn that in any restricted region of Minkowski space the Minkowski vacuum can be regarded as a KMS state having a temperature of more or less arbitrary value (depending on the normalization), typically one uses $T_{KMS}=1$. This comes about as follows. If we define the Tomita evolution by
\beq  \sigma_s(A):= \Delta^{-is}\cdot A\cdot\Delta^{is}   \eeq
with $A$ an element of the local algebra of observables under discussion, we get the KMS property in the form
\beq  (\Omega|\sigma_s(A)B\Omega)=(\Omega|B\sigma_{s+i}(A)\Omega)   \eeq

In the absence of a canonical local Hamiltonian (in RQFT the usual situation) we can equally well make the choice
\beq K:=\beta K'\quad ,\quad \Delta=:e^{-K}=:e^{-\beta K'}   \eeq
and have 
\beq \sigma_s(A)=e^{iKs}Ae^{-iKs}=e^{iK'\beta s}Ae^{-iK'\beta s}=:\sigma'_{\beta s}(A)   \eeq
We now get the following form of the KMS condition:
\beq (\Omega|\sigma'_t(A)B\Omega)=(\Omega|B\sigma'_{t+i\beta}(A)\Omega)   \eeq
Analogously we can rescale the time parameter defining
\beq \sigma'_{\beta s}:=\sigma_s\quad\text{or}\quad e^{iK'\beta s}:=e^{iKs}\quad\text{yielding}\quad \beta K'=K  \eeq
I.e., under $\sigma'_t$ we can attribute the `inverse temperature' $\beta$ to the same state $\Omega$ and we get the simple but physically remarkable result:
\begin{ob} A scale transformation of the infinitesimal generator of $\Delta^{is}$ or a rescaling of the time parameter implies a scale transformation of the temperature parameter.
\end{ob}
\begin{bem} Physically the connection between time and temperature can be understood as follows: With $t:=\beta s$ we have for the respective typical velocities in the system
\beq dx/dt=\beta^{-1} dx/ds>dx/ds\;\text{for}\;\beta<1  \eeq
that is, with regard to $t$ velocities appear to be scaled which then applies also to the respective temperature.
\end{bem}

The possible physical meaning of the local evolution given by the Tomita evolution, in particular in the generic case where we do not have a natural Hamiltonian, is discussed at length in \cite{Requ1}. It also underlies the \tit{thermal time concept} of Connes/Rovelli (see e.g. \cite{Ro1},\cite{Ro2})
We see from the above discussion that without further conceptual ingredients the notion of temperature is a little bit vague. What is in our view however certain is the underlying cause for this thermal behavior.
\begin{ob} Our detailed analysis in \cite{Requ1} shows that the primordial cause for the thermal behavior in RQFT is the underlying dynamics and pattern of the vacuum fluctuations. Their structure is exhibited depending on the different scenarios we impose.
\end{ob} 

Let us look how temperature is introduced on an advanced level in thermodynamics. 
\begin{bem} For certain reasons we avoid the way via the Carnot process which is quit common in phenomenological thermodynamics. We will nevertheless discuss Carnot processes in the relativistic regime in the following.
\end{bem}
We start from the fundamental concept of \tit{entropy}, $S$, and its \tit{maximum principle}. Furthermore we need the notion of \tit{energy}, $E$, and assume that there exists a functional dependence of entropy on energy. In the non-relativistic framework of thermodynamics we start from two systems in thermal equilbrium. We then have (with $S=S_1(E_1)+S_2(E_2)$ and $E=E_1+E_2$):
\begin{ob} From $S=max$ and $E=const$ it follows
\beq 0=dS/dE_1=dS_1/dE_1+dS_2/dE_2\cdot dE_2/dE_1=dS_1/dE_1-dS_2/dE_2   \eeq
\end{ob}
\begin{bem} This result holds also in an exterior gravitational field. Matters change if relativity theory (special or general) becomes relevant (see below).
\end{bem}
\begin{defi} We define
\beq dS/dE:=T^{-1}   \eeq
\end{defi}

Let us now assume that our state is given by a \tit{density matrix}, $\rho$,
\beq \rho=\sum_i w_i|\psi_i><\psi_i|    \eeq
with $\psi_i$ the eigenstates of some Hamiltonian.
\begin{defi} We define the v.Neumann entropy 
\beq S_{vN}:=-\sum_i w_i\ln w_i  \eeq
\end{defi}
Furthermore we have
\beq E=Tr(H\cdot\rho)=\sum_i w_iE_i   \eeq
If the $w_i$ are functionally dependent on the $E_i$, i.e. $w_i=F(E_i)$ we can, in principle, calculate the temperature, $T$, of the state. If we assume in particular that the state is an equilibrium state (maximal entropy) we get the usual Gibbs result for the temperature.
\begin{conclusion} In order to be able to define the temperature of some abstract state (as they are typically given in RQFT) we need a functional dependence of its entropy on its energy, i.e. we need some Hamilton operator (which preferably is physically motivated in some way or the other). Furthermore we need the concept of thermal equilibrium, i.e., we have to be able to compare different states of some class of our system. All this comprises the zeroth law of thermodynamics. The maximum principle of entropy then guarantees that states of different temperature equilibrate in the non-relativistic regime if brought in thermal contact.
\end{conclusion}
\begin{bem} The need to be able to bring different states in thermal contact was also emphasized in \cite{Bu}. Note however that the situation is more involved in the relativistic regime (see below).
\end{bem}
In \cite{Requ1} we derived a Tomita Hamiltonian from a density matrix, representing the local vacuum state, by writing
\beq w_i=:e^{-\lambda_i}\quad\text{and}\quad \hat{H}:=\sum_i \lambda_i|\psi_i><\psi_i|    \eeq
It is then a nontrivial task to show that such a Hamiltonian may have a deeper physical meaning.
\section{On Thermometer Observables}
We begin with a discussion of thermometer observables as introduced in \cite{Bu}. The authors follow basically the philosophy expounded in \cite{Haag} where measuring observables, called \tit{detectors}, are introduced as a subclass of the algebra of local observables. 
\begin{bem} We note that we are not entirely convinced that this choice does exhaust all the relevant cases (see below). 
\end{bem}
In any case, one should remark that real \tit{measuring operations} (usually performed with the help of quite specific devices consisting typically of many degrees of freedom (DoF)) are interferences from outside with the microscopic model system under discussion, in our case a model RQFT. That is, the detector operators introduced in \cite{Haag} are not! measuring instruments but rather certain particular observables that are measured by using external (sometimes macroscopic) instruments which do not! belong to the microscopic system. We think, this is an important distinction (see below).

In \cite{Bu} (we employ for reasons of simplicity the notation of the authors) in the case of the free massless scalar field a regularized \tit{Wick-squared} of the field $\phi(x)$ is used, i.e.
\beq \theta_0(x):=(\phi(x+z)\phi(x-z)-h(x+z,x-z))_{z=0}    \eeq
with $h$ denoting the \tit{Hadamard parametrix} (cf.\cite{Wa3}). The authors show that for KMS-states with 2-point function
\beq <\phi(x)\phi(y)>:=(2\pi)^{-3}\cdot\int dp\varepsilon(p_0)\delta(p^2)(1-e^{-\sigma p})^{-1}e^{-i(x-y)p}    \eeq
$\sigma\in\R^4$ a temperature 4-vector with $\sigma_0>|\vec{p}|$ it holds
\beq <\theta_0(x)>_{\sigma}=(12\sigma^2)^{-1}=:T    \eeq
They then argue that $\theta_0(x)$ can be used as a thermometer observable.

On the other hand, after some calculations they show that an accelerated Rindler observer (with Unruh-temperature $T=a/2\pi$) measures
\beq <\theta_a(\xi,\eta>_a\equiv 0  \eeq
in the Minkowski vacuum. As a consequence they argue that the Minkowski vacuum is perceived by an accelerated observer, carrying along with him his measuring device, as cold as for an inertial observer. We must say that this technical result does not come as a surprise to us which we are going to explain immediately.

First of all, one should note that
\begin{ob} The renormalization prescription by means of the Hadamard parametrix is the same in all the states under discussion and is the same as in the Minkowski vacuum (basically the subtraction of the expectation value of the naive 2-point function). This implies that also the thermometer observable is the same in all the states under discussion.
\end{ob}
Secondly, the Wick square is a scalar and the accelerated $:\phi^2:(\xi,\eta)$ in Rindler coordinates is the same observable $:\phi^2:(x)$ as in Minkowski coordinates at the same point.
\begin{ob} The thermometer observable used by the authors of \cite{Bu} is insensitive to acceleration.
\end{ob} 

To come to our final conclusion we now want to recapitulate the relation between the global Minkowski-vacuum theory and the corresponding restricted theory in the Rindler wedge. The wide spread standard representation of the Minkowski vacuum $|\Omega>_M$ as a vector in the Rindler tensor product $\mcal{H}_L\otimes\mcal{H}_R$ over the left and right wedge, $W_L,W_R$, in the form
\beq \label{vac} |\Omega>_M=\prod_j(N_j\sum_{n_j} e^{-\pi n_j\omega_j/a}\cdot |n_j,L>\otimes |n_j,R>)   \eeq
with $N_j=(1-e{-2\pi\omega_j/a})^{1/2}$, $\omega_j$ the energy of the Rindler modes (cf. e.g. \cite{Wa1}, reviews are \cite{Cris},\cite{Tak}) is not strictly correct. $|\Omega>_M$ has infinite norm in this representation (cf. the recent discussion in \cite{Requ1} as to the deeper reasons) but contains basically the correct physics. It shows for example immediately that its restriction to e.g. $W_R$ is a temperature (KMS) state $\rho_R$. It is evident from this construction that with $A$ a local observable in $W_R$ we have:
\beq <\Omega|A\Omega>_M=Tr(\rho_RA)  \eeq
\begin{bem} One should note that the field momentum $\pi$ is not a scalar like $\phi$. It is however shown in \cite{Fu} that nevertheless we have
\beq \partial\phi/\partial t=\partial\mcal{L}/\partial t=\partial\mcal{L}/\partial\eta=\xi^{-1}\partial\phi/\partial\eta   \eeq
\end{bem}

A rigorous discussion can be found in \cite{Kay}. From Bisognano-Wichmann (\cite{Wi}) we know that $\omega\upharpoonright_{WR}$ is a $\beta=2\pi$ KMS-state. We can however give a more explicit and intrinsic construction over the \tit{double-wedge} by employing the \tit{Fulling-quantization}, the \tit{Fulling-vacuum} being denoted by $\omega_F$. This approach is more akin to the above vector construction. With $\tilde{\omega}_F^{2\pi}$ the so-called \tit{double-wedge KMS-state} over $\mcal{H}_L\otimes\mcal{H}_R$, $\tilde{\mcal{A}}$ the corresponding algebra of observables, $\omega_0$ the Minkowski vacuum, we have
\beq \omega_0(\tilde{A})=\tilde{\omega}_F^{2\pi}(\tilde{A})  \eeq
I.e., this is exactly what has to be expected on physical grounds. Note that this was already anticipated in \cite{Fu}.
\begin{bem} As the change from Minkowski to Rindler perspective is essentially only a coordinate transformation plus a geometric restriction, fields and observables should remain the same. What only changes are the mode expansions respectively the particle concepts.
\end{bem}
These observations lead to the conclusion
\begin{conclusion} As the emergence of Rindler modes is the typical consequence of accelerated reference frames, thermometer observables which are insensitive to acceleration cannot show the desired effec (cf. also \cite{Clifton}).
\end{conclusion}

In this context we want to mention the famous \tit{clock hypothesis} (cf. e.g. \cite{Rindler} or \cite{Sexl}) which underlies already implicitly Einstein's analysis of general relativistic toy-models (employing the special relativistic framework; take for example the \tit{rotating disk model}) but was never made explicit by Einstein himself. 
\begin{ob} There exist ideal clocks which only feel the instantaneous velocity along their path but not the acceleration. They measure proper time along their orbit.
\end{ob}
\begin{bem} GR, that is, acceleration, is however implicitly present because the instantaneous inertial systems permanently change in general along the curve of the particle/clock.
\end{bem}  
Our analysis shows that most of the usual observables are also insensitive to acceleration which in our particular case is not sufficient.

Put differently, we need a measuring device which is sensitive to acceleration. We postpone a discussion of the criticism, uttered in \cite{Bu}, saying that it is not! temperature but only acceleration which is measured. In \cite{Wa2} a 2-level system is employed with states $|\uparrow>,|\downarrow>)$ (eigenstates of $H_D$)
\beq H_D=\Omega\cdot A^{\dagger}A\;,\; A|\downarrow>=0=A^{\dagger}|\uparrow>\,,\, A^{\dagger}|\downarrow>=|\uparrow>\,,\,A|\uparrow>=|\downarrow>  \eeq
which is linearly coupled to the field $\phi(\xi,\eta)$. The uncoupled measuring device is assumed to evolve according to Rindler time $\eta$ under the free evolution, induced by $H_D$. 
\begin{bem} One should note that the translation invariance under Rindler time of detectors in the Rindler scenario is different from time translation invariance in the inertial case. We think, this is an important point.
\end{bem}

We want to show how simple model Hamilton operators behave under the influence of acceleration. To keep matters from becoming unnecessarily complex we treat them non-relativistically which should be sufficient as we are only interested in qualitative results.
\begin{ob} We employ the equivalence principle and replace acceleration by a linear gravitational field. 
\end{ob}
The simplest example is the \tit{harmonic oscillator}. We include a linear term in the potential. With $a$ the acceleration we get:
\beq V(x)=cx^2+amx=c(x^2+am/c\cdot x)=c(x+am/2c)^2-(am/2c)^2  \eeq
With $y:=x+am/2c$ we see that the energy levels of the eigenstates are simply shifted by a constant $E_0=(am/2c)^2$ while the spacing remains the same. For a more general atom Hamiltonian first order perturbation theory yields
\beq E_n=E_n^{(0)}+E_n^{(1)}\quad,\quad E_n^{(1)}=\int \psi^{(0)}(amx)\psi^{(0)}dx   \eeq
That is, both the energy levels and their spacing will depend on the acceleration in the generic case.
\begin{bem} We argued in\cite{Requ1} that the observed thermal behavior of the relativistic quantum vacuum is caused by the primordial structure and dynamics of the vacuum fluctuations. Therefore it follows from our above analysis that an accelerated detector will see a different kind of vacuum fluctuation pattern compared to an inertial observer.
\end{bem}
\section{On General Relativistic Temperature}
The case of the notion of temperature in special relativity (SR) is treated in \cite{Requ2} and literature cited there. The transformation behavior of temperature in SR is also exploited in the following section about Carnot processes in the relativistic regime. While in SR these transformation properties under Lorentz boosts are (depending on the point of view) more of a formal or conceptual nature, in GR they are of a more objective observational character. As it is of direct relevance to our following discussion of the Unruh effect, we will only analyze what is called in the literature the \tit{Tolman-Ehrenfest effect} (in our view it is actually more than a mere effect).

The crucial property that makes the relativistic regime so different compared to non-relativistic thermodynamics in an exterior gravitational field is that pure \tit{heat} now has \tit{weight}! The usual set-up is an \tit{isolated} system in \tit{thermal equilibrium} in a static gravitational field. The fundamental assumption is still the \tit{entropy-maximum principle}. In a detailed but relatively special analysis (relying on the properties of a \tit{radiation thermometer}) the situation was first analyzed by Tolman/Ehrenfest (cf. \cite{Tol},\cite{Tol2},\cite{Tol3}) with the result:    
\begin{ob} In thermal equilibrium in a static gravitational field we have for an isolated system
\beq T(x)\cdot\sqrt{-g_{00}(x)}= const   \eeq
I.e., in contrast to the non-relativistic regime (cf. sect.2), there exists in general a temperature gradient in a system being in thermal equilibrium in the relativistic regime.
\end{ob}
More recent treatments can be found in\cite{Ba},\cite{Ebert} (using a general relativistic \tit{Carnot cycle} and \cite{Ro3} (relating it to the \tit{thermal time concept}). A completely different (quite sketchy) treatment is given in \cite{Landau} sect. 27. 

The entropy maximum principle allows to understand the above formula. We shall use, for reasons of simplicity, the weak field expansion of the gravitational field. With $\phi(x)$ the Newtonian gravitational potential we have
\beq \sqrt{-g_{00}}=(1+2\phi/c^2)^{1/2}\approx 1+\phi/c^2   \eeq
\begin{bem} Note that the gravitational potential is negative and is usually assumed to vanish at infinity.
\end{bem}
We assume an isolated macroscopic system to be in thermal equilibrium in such a static weak gravitational field. Its total entropy and internal energy depend on the gravitational field $\phi(x)$. We now decompose the large system into sufficiently small subsystems so that the respective thermodynamic variables can be assumed to be essentially constant in the small subsystems. As the entropy is an \tit{extensive} quantity, we can write
\beq S(\phi)=\sum_i S_i(E^0_i,V_i,N_i)   \eeq
where $E^0_i$ is the thermodynamical \tit{internal energy}, not including the respective \tit{potential energy}.
\begin{ob} It is important that in the subsystems the explicit dependence on the gravitational potential has vanished. The entropy in the subsystems depends only on the respective thermodynamical variables, the values of which are of course functions of the position of the respective subsystem in the field $\phi(x)$.
\end{ob}

At its maximum the total entropy is constant under infinitesimal redistribution of the internal energies $E^0_i$ with the total energy and the remaining thermodynamic variables kept constant. We now envisage two neighboring subsystems, denoted by (1) and (2). To be definite, we take $\phi(2)\geq \phi(1)$. We now transfer an infinitesimal amount of internal energy $dE_2^0$ from (2) to (1) (note, it consists of pure heat as for example the particle numbers remain unchanged by assumption!). As heat has weight relativistically it gains on its way an extra amount of potential energy.
\begin{ob} By transferring $dE_2^0$ from (2) to (1) we gain an additional amount of gravitational energy 
\beq dE_2^0\cdot\Delta\phi/c^2 \quad ,\quad \Delta\phi=\phi_2-\phi_1   \eeq 
It is important to realize that this gravitational energy has to be transformed from mechanical energy into heat energy and reinjected in this form into system (1) (for example by a stirring mechanism acting on system (1) and being propelled by the quasistatic fall of the energy  $dE_2^0$).
\end{ob}
The energy balance equation now reads
\beq dE_1^0=dE_2^0+dE_2^0\cdot\Delta\phi/c^2=dE_2^0(1+\Delta\phi/c^2)  \eeq 
We then have (with $dS_1=-dS_2$ in equilibrium)
\beq T_1^{-1}=dS_1/dE_1^0=-dS_2/-(dE_2^0(1+\Delta\phi/c^2))=T_2^{-1}\cdot (1+\Delta\phi/c^2)^{-1}   \eeq
that is
\begin{conclusion} It holds
\beq T_1=T_2(1+\Delta\phi/c^2)\quad ,\quad \Delta\phi:=\phi_2-\phi_1\geq 0   \eeq
\end{conclusion}
\begin{ob} The subsystem (1), having a potential energy being lower than (2), has a higher temperature.
\end{ob}

We can give the above relation another more covariant form. In the approximation we are using it holds:
\beq \frac{\sqrt{1+2\phi_2/c^2}}{\sqrt{1+2\phi_1/c^2}}=\frac{1+\phi_2/c^2}{1+\phi_1/c^2}=\frac{1+(\phi_1+\Delta\phi)/c^2}{1+\phi_1/c^2}=1+\Delta\phi/c^2   \eeq
\begin{conclusion}[Covariant form] It holds
\beq T_1\cdot\sqrt{-g_{00}(1)}=T_2\cdot\sqrt{-g_{00}(2)}=const    \eeq
\end{conclusion}
The non-infinitesimal result follows by using a sequence of infinitesimal steps.

It is perhaps useful to derive the above result in yet another, slightly different, way. In case $\phi(x)$ vanishes at infinity we can apply the entropy-maximum principle as follows. We extract the energy (bringing it to infinity)
\beq dE_2=dE_2^0+dE_2^0\cdot \phi_2/c^2   \eeq
from subsystem (2) and reinject the energy
\beq  dE_2=dE_1=dE_1^0+dE_1^0\cdot \phi_1/c^2  \eeq
from infinity into (1). We get
\beq dE_1^0=dE_2^0\cdot\frac{1+\phi_2/c^2}{1+\phi_1/c^2}  \eeq
and
\beq   T_1^{-1}=dS_1/dE_1^0=T_2^{-1}\cdot\left(\frac{1+\phi_2/c^2}{1+\phi_1/c^2}\right)^{-1}   \eeq
hence
\beq  T_1\cdot (1+\phi_1/c^2)=T_2\cdot (1+\phi_2/c^2)   \eeq

As an example one may mention the Rindler/Unruh space-time. We have in Rindler coordinates
\beq g_{00}=-\xi^2\; ,\; a=\xi^{-1}\; ,\; T=a/2\pi   \eeq
i.e.
\beq T\sqrt{-g_{00}}=2\pi^{-1}=const   \eeq
\begin{bem} Note that the Rindler/Unruh scenario is not a weak-field example. The above result shows however that the Unruh-temperature formula is consistent with the general framework, as is, for example, the Hawking effect. But  the relativistic vacuum is certainly not! a thermodynamical system in the classical sense.
\end{bem}
In the Newtonian limit potential differences are, in a sense, objective properties. In this limit the space-time is (essentially) flat. But we have:
\begin{ob} A coordinate transformation within the static regime influences the temperature function $T(x)$ while $T\sqrt{-g_{00}}$ still remains constant. We can however refer to our discussion of the connection between time and temperature in sect.2. That is, a change of metric has a natural effect on the values of the local temperatures.
\end{ob}

Another interesting question is the physical meaning of the product $T_0=T\sqrt{-g_{00}}$. There are some tentative (heuristic) remarks in e.g. \cite{Tol} and \cite{Landau}.
\begin{ob} In the Rindler/Unruh scenario $T_0$ is the local temperature along the Killing-orbit with $-g_{00}=1$.
\end{ob}
In the weak-field situation, where $\phi(x)$ is assumed to vanish at infinity and where $-g_{00}\to 1$, we can make the following thought experiment. We transport both subsystem (1) and (2) to infinity while transforming the respective gained potential energies into heat energy (as described above). We then have
\begin{ob} In the weak-field scenario the subsystems (1) and (2) will have the identical temperature $T_0$ when translated quasistatically to infinity. Reversing this process the subsystems within the gravitational field will have the correct local temperatures $T_1,T_2$ if they had the same temperature $T_0$ at infinity. In the latter process part of the internal energy has to be transformed to gravitational energy.
\end{ob}

\section{On Relativistic Carnot Processes}
In \cite{Bu} the usefulness of Carnot processes for thermodynamics in the relativistic regime is called in question, both as a way to establish an absolute temperature scale and as a means to gain work by running them between reservoirs having different temperatures. Both questions will be dealt with in greater detail in connection with the Unruh effect in the following section, which is the central section of this paper. But before doing this we want to briefly discuss the nature of Carnot processes in the relativistic regime from a more general point of view.

Carnot processes are in fact of a quite peculiar character if relativity has to be included. We treated the special-relativistic Carnot process, among other things, in some detail in sect. 4.3 of \cite{Requ2}. In order to deal adequately with this topic almost the whole body of relativistic thermodynamics has to be employed.
\begin{bem} One should be aware that various questions of principle are not really settled in this field (see the discussion and the references in\cite{Requ2}; a more up-to-date version is forthcoming). Furthermore, we restrict ourselves, for the time being, to the special-relativistic framework 
\end{bem}
It is perhaps not widely known that already v.Laue introduced the relativistic Carnot process  in his book on relativistic thermodynamics (cf. \cite{Laue}, a similar discussion ca be found in sect.70 of \cite{Tol}). Both used however the `wrong' (classical) transformation laws for temperature etc. (see the discussion in \cite{Requ2}). 

The Carnot process (in the form of v.Laue/Tolman) works as follows: A simple system (engine) operates between two reservoirs, $R_1,R_2$, with constant pressure $p=p_0$ over the full cycle. $R_1$ rests in the laboratory frame, having temperture $T=T_1$. $R_2$ moves with constant velocity $u$ , having \tit{relativistic temperature}
\beq T_2=\gamma\cdot T_1\quad\text{with}\quad \gamma=(1-u^2/c^2)^{-1/2}\quad\text{the Lorentz factor}   \eeq
\begin{bem} i) Such a process at constant pressure can be realized by a system consisting of a coexisting fluid and gaseous phase.\\
ii) Note that both v.Laue and Tolman belong to the classical period, using a moving temperature
\beq T_2=\gamma^{-1}\cdot T_1   \eeq
Einsteins opinion in this context is particularly interesting, cf. \cite{Liu} and \cite{Requ2}. One should however emphasize that the transformation properties of temperature are particularly disputable.\\
iii) All the processes in the following are performed quasistatically as usual in thermodynamics. 
\end{bem}

The system is initially at rest, having internal energy $E_a$, volume $V_a$, temperature $T_1$. It absorbs the amount of heat $Q_1$ from $R_1$ and does the work $p_0(V_b-V_a)$ on the environment. We then have (first law of thermodynamics):
\beq Q_1=(E_b-E_a)+p_0(V_b-V_a)=H_b-H_a \quad\text{(enthalpy)}   \eeq
In the second step the system is adiabatically accelerated to the velocity $u$ of $R_2$. The work done by the system is
\beq W_2=E_b-E_c\quad\text{with}\quad E_c=(E_b+p_0V_b\cdot u^2/c^2)\cdot\gamma    \eeq
(cf. e.g. \cite{Requ2}). In the third step the amount of heat being released to $R_2$ is $Q_2$ and the amount of work done is $W_3=p_0(V_d-V_c)$. We assume now that the amount of released heat in step (3) is just sufficient so that the system returns to its initial state by a reversible deceleration. This is exactly the case if
\beq Q_{2.0}=(E_d-E_c)_0+p_0(V_d-V_c)_0=-Q_{1,0}=Q_1   \eeq
with the subscript `0' denoting the proper values of the respective quantities. That is, $Q_{2.0}$ is the amount of heat measured by a comoving observer. The work done by the system in this last step is $W_4=E_d-E_a$.

The first law of thermodynamics tells us that 
\beq Q_1+Q_2=W_1+W_2+W_3+W_4   \eeq
\begin{ob} We have (cf. \cite{Requ2})
\beq Q_2=-Q_1\cdot\gamma\quad\text{and}\quad \Delta W=Q_1\cdot (1-\gamma)<0      \eeq
\end{ob}
That is, in each cycle there is a heat transport from reservoir $R_1$ to $R_2$ and a negative amount of work done by the system on the environment (if the engine is operated in this direction).
\begin{bem} Note that quantities like $Q_2=-Q_1\cdot\gamma$ and $p_0(V_d-V_c)$ are relativistic values as `observed' in the laboratory frame. We remarked already above that there is no uniform opinion as to there ontological status. By running the cycle in the opposite direction we of course gain positive work.
\end{bem}

The following observation is important for the understanding of the thermodynamics of the Unruh effect. We have
\beq Q_{2,0}=-Q_{1,0}=Q_1\quad\text{and}\quad V_c=\gamma V_b,V_d=\gamma V_a   \eeq
i.e.
\beq p_0(V_d-V_c)=\gamma\cdot p_0(V_b-V_a)   \eeq
\begin{conclusion} If one adopts the point of view that these relativistic transformation laws are rather employed for reasons of consistency and conceptual coherence in the relativistic domain one observes that both
\beq Q_{2,0}=-Q_{1,0}\quad\text{and}\quad W_{3,0}=-W_{1,0}   \eeq
I.e., from this point of view the negative work $\Delta W$, got out of one cycle, consists of the surplus of  $|W_2|$ compared to $W_4$. The system at the end of process (1) has gained surplus energy (i.e. relativistic mass) which then has to be brought to velocity $u$. In the process of deceleration the system has less weight, thus less work is gained.
\end{conclusion}
Furthermore we calculated the \tit{Carnot efficiency} in \cite{Requ2}
\begin{koro} The Carnot efficiency is
\beq \eta=1-Q_1/Q_2=1-T_1/T_2=1-\gamma^{-1}>0  \eeq
\end{koro}

In section 5 of \cite{Requ2} we then showed that an exchange of heat between $R_1,R_2$ does not change the total entropy which means:
\begin{ob} The two subsystems (reservoirs) $R_1.R_2$ are in thermal equilibrium if 
\beq T_2=T_{2,0}\cdot\gamma=T_1\cdot\gamma=T_{1,0}\cdot\gamma  \eeq
That is, the two rest temperatures have to be equal while the moving temperatures are different!
\end{ob}
We shall see in the next section that a closely related situation prevails in the Unruh effect if one tries to establish a similar cycle. I.e., the thermal energy we get out of the system is part of the mechanical work of acceleration of the probe.   
\section{On the Reality of Unruh Temperature}
After having analyzed the temperature concept in varying degrees of generality in GR we want to report in this concluding section, in the light of our previos observations, on the detailed analysis of acceleration thermodynamics done by Unruh/Wald in \cite{Wa1},\cite{Wa2}. In \cite{Wa1} the interesting concept is \tit{flat-spacetime mining} which the authors compare with the corresponding phenomenon in BH-thermodynamics. In both papers the authors are particularly concerned with the complementary viepoints of the accelerated and inertial observer (cf. our own remarks in the introduction). It is fascinating that one gets in the end the same result, while the technical details and concepts are quite different in the two pictures. While \cite{Wa1} deals primarily with the thermodynamical and radiation aspects of the Unruh effect, \cite{Wa2} discusses the behavior of accelerated detectors. We will begin with the discussion of acceleration thermodynamics and what the authors call \tit{mining of flat space}.
\begin{bem} As the quite intricate technical calculations can be found in \cite{Wa1} we discuss in the following mainly the principal physical aspects of the subject matter.
\end{bem}

Our primary concern will be the following. As the reality of Unruh temperature as a thermodynamical property beyond some vague formal meaning is called into question by some authors, we try to show that one can really extract thermodynamical heat energy, carried by some thermodynamical system, from the acceleration regime and transport it to our laboratory where it can then be manipulated in the usual thermodynamic way. We will use (cf. \cite{Wa1}) a box with perfect reflecting walls having a door which can be opened and closed. We then perform the following mining process:\\
i) The box is started from rest and is slowly brought to constant acceleration $a$. The box undergoes \tit{rigid} acceleration in, say, the x-direction (cf. for example \cite{Rindler}). This implies that the front and back face have slightly different proper acceleration, each moving along its own orbit. We assume that the box is small so that this difference is not to large (for calculational reasons).\\
ii) At constant acceleration $a$ the door is opened allowing the box to fill with thermal (Unruh) radiation (accelerated observer viewpoint!) and then closed again.\\
iii) The box is slowly decelerated and brought back to rest in the laboratory or allowed to fly off with constant velocity.
\begin{ob} We shall find that the box is filled with thermal radiation at temperature $T=a/2\pi$.
\end{ob}
  
Our strategy is now the following. We can be convinced that the \tit{accelerated observer} viewpoint , that is, that he sees thermal \tit{accelerated radiation}, is an objective real phenomenon, if we are able to describe the whole process from the viewpoint of an \tit{inertial observer}. Here comes into play the surprising behavior of \tit{moving mirrors} (cf. \cite{Dav},\cite{Dav1},\cite{Dav2},\cite{Moore},\cite{Helfer},\cite{Alves}, to mention just a few old and more recent sources). 
\begin{bem} While perhaps surprising, the mechanism at work in the moving mirror situation is similar to the Unruh-effect. In the mirror case we have in effect a \tit{time dependent} (particle creating) background. For certain mirror trajectories we even observe a thermal spectrum.
\end{bem}
\begin{ob}[Moving Mirror] It can be shown that mirrors which undergo a changing acceleration radiate. In particular, in the direction of increasing acceleration there is a negative flow of energy while in the direction of decreasing acceleration there is a positive energy flow. In the opposite directions we have the opposite signs. If the mirror is asymptotically static after a period of increasing acceleration and then decreasing acceleration the net energy flow is positive in the direction of acceleration. In the opposite direction it is negative.
\end{ob}
\begin{bem} Note that in the period of increasing acceleration the rear end of the above box has a larger proper acceleration than the front end which has an effect on the interior of the box!
\end{bem}

We now come to the following conclusion. 
\begin{conclusion} [Inertial observer viewpoint] i) During step i) (quasistatical increase of acceleration of the empty closed box) the interior remains in its ground state, energy flows out of the box and the field energy in the box becomes negative. The ground state in the box is the so-called Rindler/Fulling vacuum which is energetically lower than the Minkowski vacuum (see \cite{Fu}).\\
ii) In step ii) the negative energy is radiated out of the open box and the state of the box becomes the Minkowski vacuum.\\
iii) During deceleration a positive energy flux enters the box (created by the accelerating rear and front mirror). The box arrives in the laboratory filled with radiation of positive energy density and temperature $a/2\pi$.
\end{conclusion}
\begin{bem} The total energy radiated to infinity in this process is (and has to be) positive as the vacuum is a state of minimum energy (as to a general theorem cf. \cite{Worono}, concrete calculations can be found in \cite{Wa1}).
\end{bem}

What have we learned from the study of the above thermodynamical cycle? First, we have seen that, while accelerated and inertial observer describe the various steps quite differently, they come nevertheless to the same final conclusion. That is, thermal radiation (having the expected Unruh temperature) has been extracted from the vacuum and has been brought to the laboratory. Second, what is the energetic source fuelling this process? 
\begin{ob} The original energy source is the work done by the external accelerating agent. That is, the possibility to extract energy from the vacuum is actually no mystery!
\end{ob}   
\begin{bem} We want to emphasize the similarity to the special-relativistic Carnot process we have discussed in the preceding section. There the work gained stemmed also from accelerating certain heat quantities.
\end{bem}

We want to conclude this section with a brief discussion of the surprising behavior of a Rindler particle detector from the viepoint of an inertial observer (\cite{Wa2}). While the accelerating observer registers the absorbtion of a Rindler quantum the inertial interpretation is quite surprising at first glance. 
\begin{ob} The inertial description of the detectors reaction is the following. The detector absorbs part of the existing vacuum fluctuations in the region (e.g. the right wedge $W_R$) where the detctor is located while in the non-causally related left wedge $W_L$ a real Minkowski particle is emited, the reason being the strong translocal entanglement between right and left wedge in the Minkowski vacuum (see formula (\ref{vac}) in sect.3).
\end{ob}
\begin{bem} This translocal coupling between causally non-related regions has been discussed in quite some detail in \cite{Requ1}. It is closely connected with the so-called Tomita-Takesaki theory in the field of v.Neumann algebras.
\end{bem}  
\section{Conclusion}
We have shown that, whereas the situation is considerably more intricate in the (general) relativistic regime, most of the known thermodynamic processes have their counterpart in GR. One can make an apriori definition of \tit{absolute temperature} via the \tit{entropy-energy relation}, thus avoiding to start from (generalized) Carnot processes, which, however, can in principle also be done in GR. Furthermore, in order to measure the so-called \tit{Unruh-temperature}, we have to employ observables/detectors which are sensitive to \tit{proper acceleration}.

With the help of a \tit{mining process}, described in \cite{Wa1}, we can extract \tit{acceleration radiation} from the vacuum and transfer it to the laboratory and thus are able to show that the\tit{Unruh/Rindler quanta} are real. It should however be noted that the inertial observer viepoint is quite distinct from the accelerated observer description of the respective processes. Correspondingly, the inertial desciption of the reaction of a Rindler-particle detector to Rindler quanta (accelerated observer viepoint) is surprising at first glance (cf. \cite{Wa2}) and again exhibits the relativity of the particle concept. 
\begin{ob} We emphasize that what actually is shown is the reality of vacuum fluctuations and their coherent and strongly entangled pattern permeating (or, rather, making up) space-time (\cite{Requ1}). 
\end{ob}

To add some minor points, we have shown that the results derived in the Unruh-scenario fit very well into the general thermodynamic framework in GR described for example by the Tolman-Ehrenfest effect. Furthermore we mention the simple but nevertheless important relation between time- and temperature concept in this field (underlying also the Unruh-effect). A last point to mention is the question of the degree of generality of our analysis. The analysis in e.g. \cite{Wa1} and \cite{Wa2} is primarily made for \tit{radiation}, that means mass zero quanta. In the moving mirror calculations certain boundary conditions were used which are typical of (quantum) electrodynamics. Another point is the habit to treat only vacua of particular model field theories. The true physical quantum vacuum however is the ground state of all possible fields. I.e., there should occur a whole bunch of Unruh/Rindler excitations at the same Unruh-temperature.

\end{document}